%% file: ISMIR_template.tex
\documentclass{article}
\usepackage[T1]{fontenc} 
\usepackage[utf8]{inputenc} 
\usepackage{ismir,amsmath,cite,url}
\usepackage{graphicx}
\usepackage{color}
\usepackage{booktabs}
\usepackage{tabularray}
\usepackage{multirow}
\usepackage{amsmath}
\usepackage{amssymb}
\usepackage{array}
\usepackage{makecell}
\usepackage{lineno}
\usepackage{xcolor}
\usepackage{multirow}

\title{GraphMuse: A Library for \\ Symbolic Music Graph Processing}

\oneauthor
  {Emmanouil Karystinaios$^1$ \hspace{1.5cm} Gerhard Widmer$^{1,2}$} {$^1$ Institute of Computational Perception, Johannes Kepler University Linz, Austria \\ $^2$ LIT AI Lab, Linz Institute of Technology, Austria\\ {\tt firstname.lastname@jku.at}}

\def\authorname{E. Karystinaios and G. Widmer }


\begin{document}

\maketitle

\begin{abstract}


Graph Neural Networks (GNNs) have recently gained traction in symbolic music tasks, yet a lack of a unified framework impedes progress. Addressing this gap, we present GraphMuse, a graph processing framework and library that facilitates efficient music graph processing and GNN training for symbolic music tasks. Central to our contribution is a new neighbor sampling technique specifically targeted toward meaningful behavior in musical scores. Additionally, GraphMuse integrates hierarchical modeling elements that augment the expressivity and capabilities of graph networks for musical tasks. Experiments with two specific musical prediction tasks -- pitch spelling and cadence detection -- demonstrate significant performance improvement over previous methods. Our hope is that GraphMuse will lead to a boost in, and standardization of, symbolic music processing based on graph representations. The library is available at  \url{https://github.com/manoskary/graphmuse}

\end{abstract}

\section{Introduction}\label{sec:introduction}

Symbolic music processing entails the manipulation of digital music scores, encompassing various formats such as MusicXML, MEI, Humdrum, **kern, and MIDI. Unlike audio-based representations, symbolic formats offer granular information on note elements, including onset, pitch, duration, and other musical attributes like bars and time signatures.

While prior research in symbolic music processing often adopted techniques from the image processing~\cite{Benetos2012Score-informedTutoring,Velarde2016ComposerPiano-rolls,lopez2021augmentednet} or natural language processing~\cite{miditok2021,Rutte2022FigaroControl,Liu2022SymphonyModel} domains, recent attention has shifted towards graph-based models, which could presumably better capture the dual sequential and hierarchical nature of music. Graph Neural Networks (GNNs) have been showcased as potent tools for diverse symbolic music tasks, including cadence detection\cite{karystinaios2022cadence}, optical music recognition~\cite{baro2022musigraph}, music generation~\cite{graphgen}, Roman numeral analysis~\cite{karystinaios2023roman}, composer classification~\cite{zhang2023symbolic}, voice separation\cite{karystinaios2023voice}, and expressive performance rendering\cite{jeong2019graph}. However, a standardized framework for constructing and processing music graphs has not yet been introduced to the field. To address this challenge, we developed GraphMuse, a Python-based framework to efficiently and effectively process information from musical scores, construct musically meaningful graphs, and facilitate the training of graph-based models for symbolic music tasks.

A key innovation of our work lies in the introduction of a new
sampling technique tailored to specific properties of music while maintaining efficient and robust training of GNNs. 
Additionally, GraphMuse integrates within the graphs and models hierarchical elements that augment the capabilities of graph networks for musical tasks. 

We evaluate our framework on pitch spelling and cadence detection tasks, comparing it against existing state-of-the-art methods. Through the synergistic utilization of our framework's components, we achieve a significant performance increase compared to the previous methods. Our overarching objective is to establish a standardized framework for graph processing in symbolic music analysis, thus catalyzing further progress in the field.

Altogether, our contributions are three-fold: i) We provide a structured, generic, and flexible framework for graph-based music processing; ii) we release an open source \textit{Python} library that comes with it; iii) we achieve performance improvements in a principled way by focusing on the design of the individual parts of the framework.


\section{Processing Music Scores with GNNs}\label{sec:related}


\begin{figure*}[tbp]
\centerline{\includegraphics[width=\textwidth]{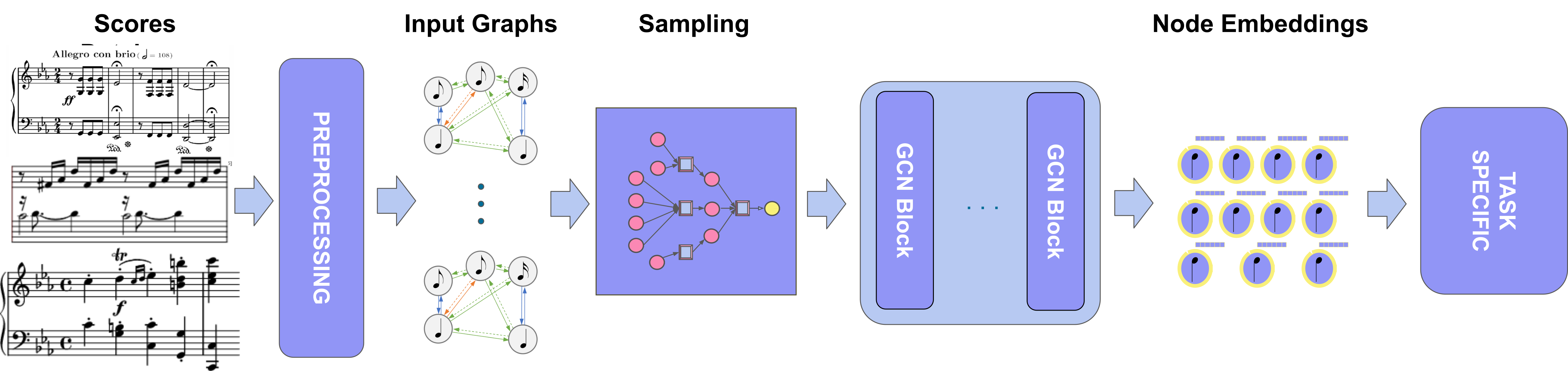}}
\caption{The general graph processing/training pipeline for symbolic music scores involves several steps: i) Preprocess the database of scores to generate input graphs; ii) Sample the input graphs to create memory-efficient batches; iii) Form a batch as a new graph with nodes and edges from various input graphs; iv) Sample a subset of nodes (target nodes) and their neighbors from the input graphs; v) Update the target nodes' representations through graph convolution to create node embeddings; vi) Use these embeddings for task-specific applications. Note that target nodes may include all or a subset of batch nodes depending on the sampling strategy.}
\label{fig:general}
\end{figure*}

In this section, we describe existing graph modeling approaches for musical scores. They all have a common pipeline which involves building a graph from a given musical score (see Figure~\ref{fig:general}) and using a series of convolutional blocks to produce context-aware hidden representations for each node. We start by describing the graph-building procedure and a generic graph convolutional block; we then take a detailed look at the problem of graph sampling, which will motivate a new sampling procedure that will be presented in the next section.

\subsection{Preprocessing: Constructing Graphs from Scores}\label{subsec:graph_creation}

A score graph can be represented as a heterogenous attributed graph. A heterogeneous graph has a type associated with each node and edge in the graph~\cite{wang2019heterogeneous}. An attributed graph has an associated feature vector for each node in the graph~\cite{gnnsurvey_tnnls21}. Therefore, a heterogenous attributed graph is defined by a quintuple $G = (V, E, X, \mathcal{A}, \mathcal{R})$, together with the mappings $\phi : V \to \mathcal{A}$ and $\psi : E \to \mathcal{R}$, where $V$ is the set of nodes, $E$ is the set of edges, $X\in V\times \mathbb{R}^k$ the feature matrix $\mathcal{A}$ is the node types and $ \mathcal{R}$ is the edge types. $\phi$ maps each node to its type and $\psi$ maps its each edge to its corresponding type.

We create such a graph from a musical score by following previous work~\cite{karystinaios2023voice,karystinaios2023roman, jeong2019graph,zhang2023symbolic}.
Each node $v \in V$ corresponds to one and only one note in the musical score. $\mathcal{R}$ includes 4 types of relations: onset, during, follow, and silence, corresponding, respectively, to two notes starting at the same time, a note starting while the other is sounding, a note starting when the other ends, and a note starting after a time when no note is sounding. The inverse edges for during, follows, and silence relations are also created. 

Formally, let us consider three functions $on(v)$, $dur(v)$, and $pitch(v)$ defined on a note $v\in V$ that extract the onset time, duration, and pitch, respectively. A typed edge $(u, r, v)$ of type $r\in\mathcal{R}$ between two notes $u, v \in V$ belongs to $E$ if the following conditions are met:
\begin{itemize}
    \item $on(u) = on(v) \rightarrow{} r = \textrm{onset}$ 
    \item $on(u) > on(v) \land on(u) \leq on(v)+dur(v) \rightarrow{} r = \textrm{during}$ 
    \item $on(u) + dur(u) = on(v) \rightarrow{} r = \textrm{follow}$
    \item $on(u) + dur(u) < on(v) \land \nexists v' \in V, \; on(v') < on(v) \land on(v') > on(u) + dur(u) \rightarrow{} r = \textrm{silence}$
\end{itemize}
$\mathcal{A}$ in the literature usually only includes a single type, i.e. the note type $\nu$. However, we extend this definition in Section~\ref{subsec:preprocessing}.

\subsection{Encoding: Graph Convolution}\label{subsec:convolution}

Graph convolution and message passing are core operations in graph neural networks (GNNs) for learning node representations. In graph convolution, in its simplest form, each node aggregates messages from its immediate neighbors by computing a weighted sum of their features:

\begin{equation}
    \mathbf{h}_v^{(l+1)} = \sigma \left( \left(\sum_{u \in \mathcal{N}(v)} \mathbf{W}^{(l)} \mathbf{h}_u^{(l)} \right) + \mathbf{h}_v^{(l)}\right)
\end{equation}

where \(\mathbf{h}_v^{(l)}\) is the representation of node \(v\) at layer \(l\), \(\mathcal{N}(v)\) denotes the neighbors of node \(v\), \(\mathbf{W}^{(l)}\) is  a learnable weight, and \(\sigma\) is a non-linear activation function. Through successive iterations of message passing and aggregation, each node refines its representation by incorporating information from increasingly distant nodes in the graph, ultimately enabling the network to capture complex relational patterns and dependencies within the graph data.

In the context of music, graph convolution can be understood as a method for defining a note not only by its own characteristics and properties but by also considering the characteristics of its neighboring notes within the musical graph. In this work, as well as previous graph-based work on music~\cite{karystinaios2022cadence,karystinaios2023roman,zhang2023symbolic} the preferred graph convolutional block is \textit{SageConv} taken from one of the first and fundamental works in graph deep learning~\cite{hamilton2017inductive}.

\subsection{Sampling: Handling Graph Data for Training}\label{subsec:rel_sampling}

In an ideal world without computing resource considerations, we can imagine a training pipeline that receives an entire graph as input to a graph convolutional model. Assuming that we have the resources and time to perform such a task the process is easy to grasp. All nodes of the graph are updated in a single step based on their neighbors as described in the previous section.

However, the graph world presents us with several complexity issues.
Graph datasets in the wild typically come in two forms: i) a
(possibly large) collection of small graphs, each containing maybe fewer than 50 nodes\cite{gnnsurvey_tnnls21}; ii) a single large-scale graph such as a social network~\cite{kipf2016semi}, a recommender system~\cite{ying2018graph}, or a knowledge graph~\cite{schlichtkrull2018modeling}. The previous naive scenario presents a time-efficiency and computation waste bottleneck for the former and a memory insufficiency issue for the latter. To mitigate these issues, in the former case one can batch many small graphs together to maximize the available resources and reduce the computation time, then the full graphs can be updated during convolution within each batch. 

\begin{figure}[btp]
    \centering
    \includegraphics[width=0.8\columnwidth]{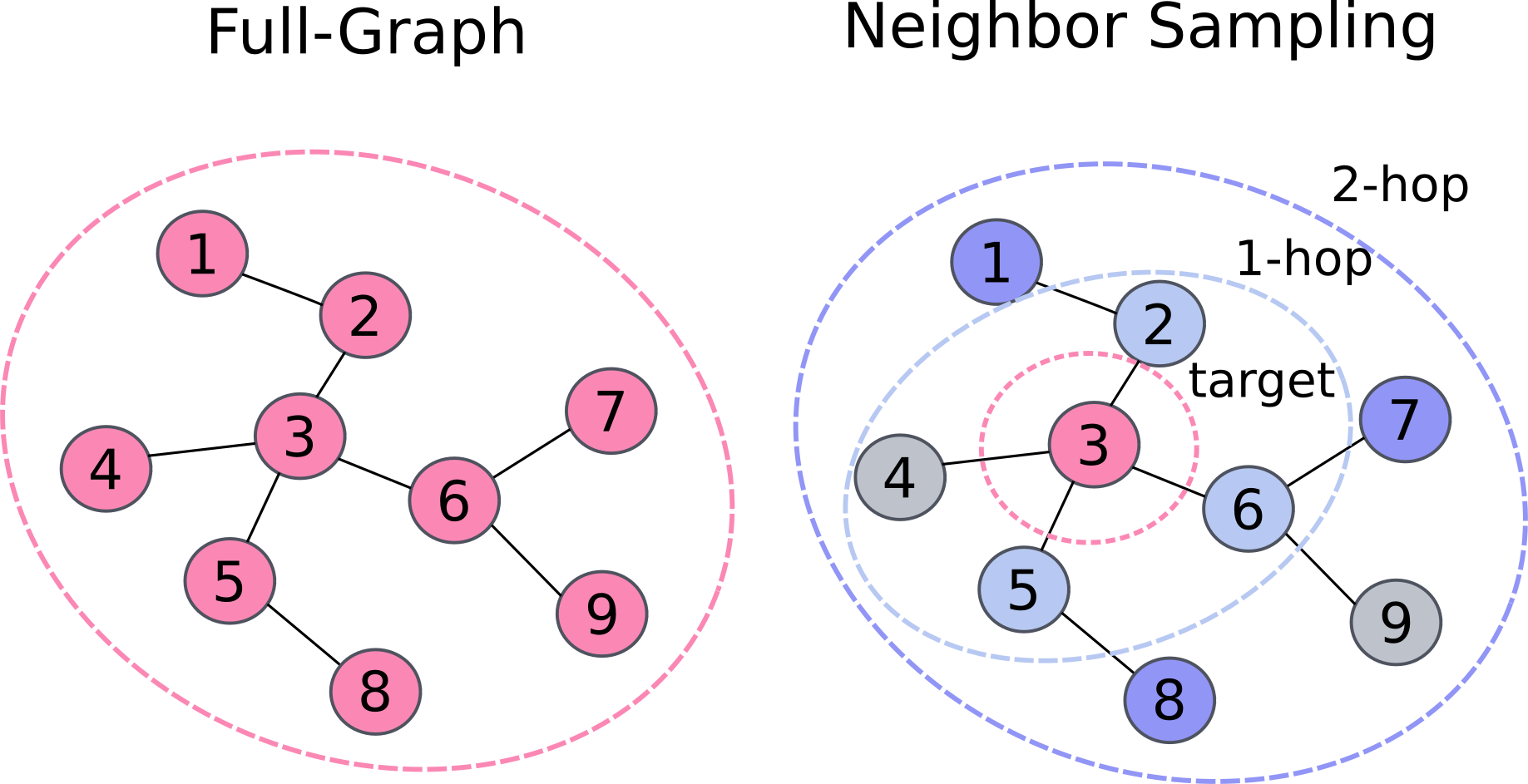}
    \caption{Full graph vs neighbor sampling. The pink-colored nodes are selected for convolution by message passing. With neighbor sampling, the pink node is the one whose representation is ultimately updated after convolution (however, for the blue nodes also take part in the convolution process as its context).}
    \label{fig:nodewise}
\end{figure}

Training Graph Convolutional Networks (GCNs) for large-scale graphs is a bit more complicated. Such graphs can be exceptionally large -- for example, the 2019 Facebook social network boasted 3.51 billion users\footnote{https://zephoria.com/top-15-valuable-facebook-statistics}. To train models with such graphs we need to devise a sampling algorithm to derive subgraphs in steps~\cite{hamilton2017inductive, zou2019ladies,zeng2019graphsaint,liu2021sampling_surver}. Such an algorithm may, for example, choose a subset of nodes across the graph and perform random walks to fetch a subset of the k-hop neighbors for the sampled nodes~\cite{hamilton2017inductive}. This process, called \textit{neighbor sampling} or \textit{node-wise sampling}, is shown in Figure~\ref{fig:nodewise} and compared to the full-graph process. 

Musical score graphs fall in between the two 
scenarios, varying notably in size. For instance, a Bach Chorale might contain 100 notes, while a Beethoven Sonata could exceed 5000 notes, with each note corresponding to a graph node. Furthermore, a musical dataset may contain many such graphs. Therefore the question arises how to efficiently train models on music graph datasets.

Since music graphs are not uniform enough to be batched together like small graph datasets, we investigate the suitability of neighbor sampling methods for music graph processing, taking into account special properties relevant in music. Standard neighborhood sampling would sample notes across different scores and fetch neighbors for those notes, creating a subgraph that can maximize the use of the available resources during training.

However, music has a specific coherence, in both the horizontal (time) and vertical (chords, harmonies) dimensions, which makes sampling approaches from the literature\cite{liu2021sampling_surver} not appropriate for music. Specifically, sampling and updating/encoding single notes without simultaneously doing so also to notes in their local context makes it difficult to learn properties that persist in time (such as local key or a harmonic function).
In this work, we address this issue by presenting a simple and musically intuitive sampling process for graphs that efficiently creates batches containing musically related notes
which, as experiments will show, can notably improve the learning results.



\subsection{Task-specific Modeling}

Finally, the node embeddings created by the graph convolutional encoder serve as input to task-specific models that solve some specific prediction or recognition task. In a graph context, we distinguish, at an abstract level, between node classification, link prediction, and entire graph classification tasks.
Examples of node classification tasks can be found in \cite{karystinaios2022cadence} which takes the embeddings from the GCN encoder and employs an edge decoder coupled with a graph convolution classifier for cadence prediction labels; and in \cite{karystinaios2023roman}, which forwards the embeddings to sequential layer and then MLP classifiers to perform Roman Numeral Analysis.
In \cite{karystinaios2023voice}, musical voice separation is framed as a link prediction task; the node embeddings are input to a pairwise edge similarity encoder to predict link probabilities between notes in the same voice.
An example of a graph classification task can be found in \cite{zhang2023symbolic} where the embeddings are aggregated and passed through a shallow MLP for composer classification.

Naturally, task-specific models will not be part of the generic graph processing pipeline and library which we publish with this paper.



\section{Methodology}\label{sec:Methodology}

\begin{figure}[t]
    \centering
    \includegraphics[width=\columnwidth]{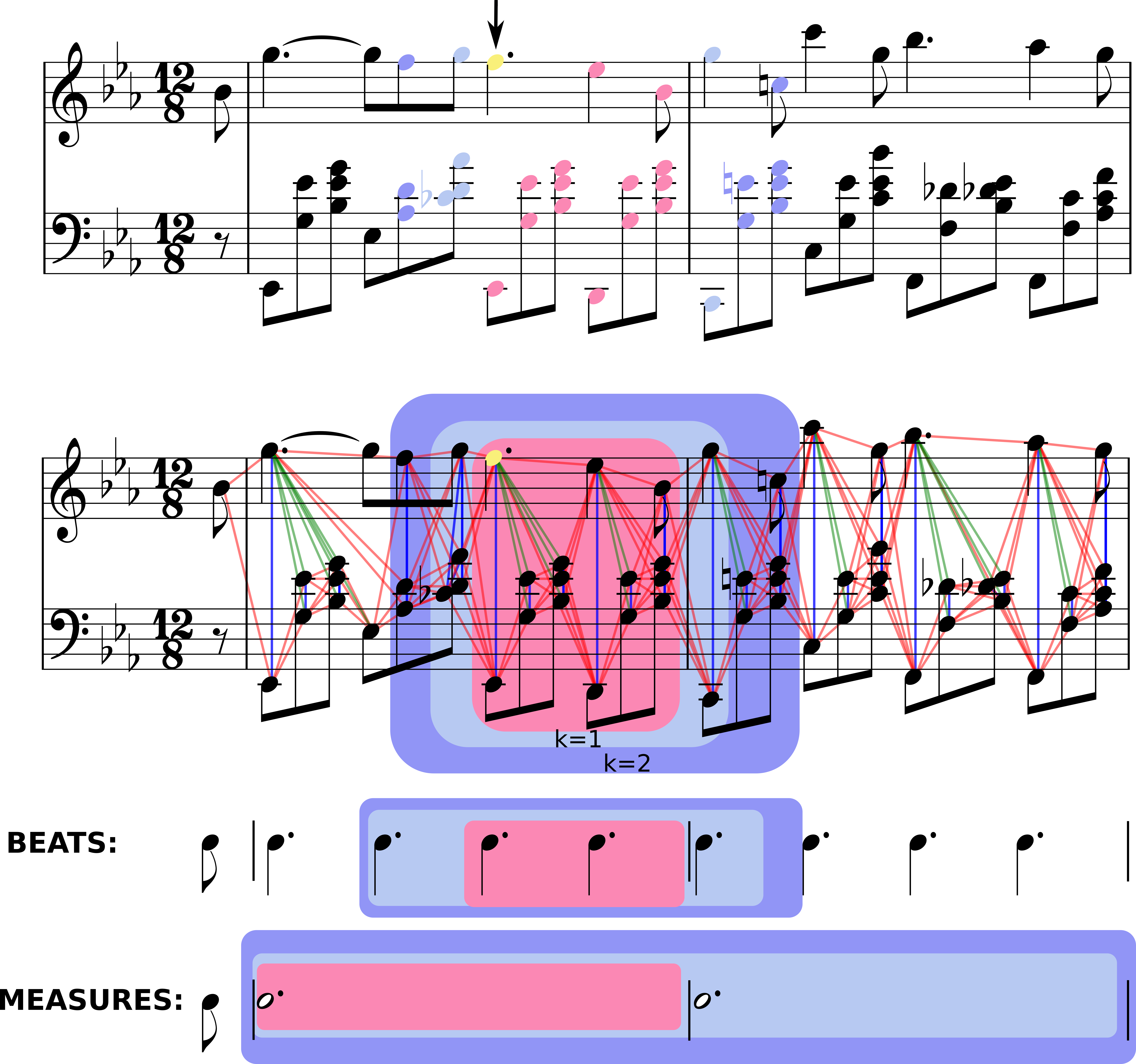}
    \caption{Sampling process per score. Top: sampled notes and their neighbors; middle: score graph and sampling process; bottom: sampling process for beats and measures.
    A randomly selected note (in yellow) is first sampled. The boundaries of the target notes are then computed with a budget of 15 notes in this example (pink and yellow notes). Then the k-hop neighbors are fetched for the targets (light blue for 1-hop and darker blue for 2-hop). The k-hop neighbors are computed with respect to the input graph (depicted with colored edges connecting noteheads in the figure). We can also extend the sampling process for the beat and measure elements (introduced in Section~\ref{subsec:preprocessing}). 
    Note that the k-hop neighbors need not be strictly related to a time window.  
    }
    \label{fig:sampling}
\end{figure}

In this section, we discuss our approach to addressing the different components of the pipeline shown in Figure~\ref{fig:general}. In particular, we explain the preprocessing procedure for creating score graphs, we detail our strategy for musically intuitive graph sampling, and finally, we discuss model variants that are made possible by the previous steps of the pipeline.

\subsection{Preprocessing}\label{subsec:preprocessing}


The central activity in the preprocessing step is the creation of graphs from musical scores. In our library, we extend the conventional graph creation process by introducing hierarchical musical dimensions (beats and measures), in order to enhance the score graphs' representational capacity. More specifically, we enrich the node type set
$\mathcal{A}$ (defined in Section~\ref{subsec:graph_creation}) with two additional types 
$\beta$ and $\mu$ for beats and measures respectively.
The process involves detecting beats and measures within the musical score, generating edges (of type  \textit{connect} to every beat from each note falling within its temporal boundaries, and repeating this process for measures. Additional edges of type \textit{next} are drawn between consecutive beats and measures to enrich the connectivity and contextual understanding within the graph. Furthermore, we aggregate features from constituent notes through the \textit{connect} edges via message passing to equip each beat and measure with informative attributes by computing the mean vector of their note features. 

The inclusion of beat and measure node elements, as well as the creation of inverse edges, are made optional, ensuring compatibility with diverse research needs and avoiding imposing rigid structures onto the graph-based music processing framework.

We prioritize the efficiency and speed of the graph creation process by transitioning the graph creation implementation to C code, leveraging its performance benefits, and establishing a Python binding for seamless integration into our workflow. Recognizing the temporal nature of musical elements, such as notes, beats, and measures, we refine our neighbor searching windows accordingly, optimizing computational efficiency.


\subsection{Sampling}\label{subsec:sampling}

We discussed general neighbor sampling for large-scale graphs in Section~\ref{subsec:rel_sampling} and some problems related to graph-structured music data. In this section, we elaborate on our musically informed sampling process for music graphs, which enables the training of the models outlined in the subsequent sections. In this process, we aim to sample sections of scores and employ neighbor sampling to fetch the neighbors of notes within those sections.

Indeed while our nodes could be ordered in various ways, the most perceptually significant aspect is time organization. Recognizably, individuals can still identify a musical piece when segmented along the time axis, whereas focusing solely on pitch intervals may be challenging. Moreover, perceptual research indicates that the commencement time of a note holds greater salience than its offset time, particularly for percussive instruments like the piano, where the sound naturally fades over time~\cite{klapuri}. Hence, when constructing graphs from musical scores, we prioritize node arrangement based on absolute onset time followed by pitch.

Our initial limitations are mostly related to memory usage. To limit our memory we need to predefine three initial arguments: i) the size of each target subgraph $S$ from every score, ii) the number $B$ of scores in each batch, and iii) the number of hops and neighbors for each hop (similar to node-wise sampling techniques). In each batch, we update the representation of our target nodes which is essentially the size of $S \times B$.

Once the ordering is set and the three arguments are defined we can initiate the process of sampling a subgraph, as shown in Figure~\ref{fig:sampling}. First, we sample a random note from the graph of each score. Next, we correct the position of the note by searching for any vertical neighbors (same onset value notes and potentially different pitch). Then we extend to $S$ notes to the right where $S$ indicates a predefined maximum subgraph size. We also correct the rightmost boundaries to include or exclude vertical neighbors for the last onset always respecting the aforementioned size $S$. Once this process is completed we obtain the target nodes per score within the batch. These are the nodes whose representation we want to update at the end of the graph convolutional process.

However, since graph convolution is performed recursively we need to fetch the $k$-hop neighbors for each one of the target nodes where $k$ indicates the depth of the GCN. For this step, we can consult the literature~\cite{hamilton2017inductive} and perform neighbor sampling to fetch the $k$-hop neighbors. This process is repeated for $B$ different scores. Finally, the $B$ score subgraphs of size at most $S$ each are first joined together and then fed to the model. 

During this process, we can keep information about the target nodes and the size of each score subgraph, which could allow us to design more creative models that can exploit this information. Such models are presented in the next section. Moreover, we adopt a potential approach for hierarchical graphs by also extending the sampling for beat and measure nodes as shown in Figure~\ref{fig:sampling}.


\subsection{Model Designs}\label{subsec:models}


In this section, we explore various model designs for the graph-based encoder in our processing pipeline (Figure~\ref{fig:general}). Designing such an encoder involves addressing two fundamental questions: the selection of graph convolutional blocks and the selective exploitation of information from the input graph.

The first question, regarding graph convolutional blocks, remains open-ended, offering numerous possibilities for exploration and customization. In its current version, \textit{GraphMuse} offers the options of convolutional blocks on a per-node or per-edge type basis. We suggest that graph-attention networks may offer promising avenues, particularly for hierarchical elements such as beats or measures.

In response to the second question, we devise a series of models by selectively incorporating or excluding elements from the input graph. Our foundational model, termed \textit{NoteGNN}, exclusively utilizes note elements and their corresponding edges. This model serves as the basis for further extension. For instance, we expand upon \textit{NoteGNN} to construct \textit{BeatGNN}, which incorporates beat elements (see Section~\ref{subsec:preprocessing} above) alongside notes. Similarly, we develop \textit{MeasureGNN} by integrating measures into the encoding process. When all note, measure, and beat elements are included, the resultant model is denoted as \textit{MetricalGNN}. 

Furthermore, we explore the possibility of hybridizing model types, such as combining GNNs with sequential models. This hybridization is facilitated by the sampling process that organizes notes in onset order, allowing for the batch to be unfolded by score. Consequently, the same batch can be processed through both GNN and sequential models simultaneously.
Specifically, we employ a graph encoder and a sequential encoder in parallel -- in our case we use a stack of 2 bidirectional GRU layers. The GRU stack receives the unfolded batch of size $(B, S, K)$ where $B$ is the number of scores within the batch, $S$ is the number of sampled target nodes for each score order by onset and then by pitch, and $K$ is the number of node features. The embeddings of both encoders are concatenated together and an additional linear layer is applied to project them to the required dimension.

This architecture, which we call \textit{HybridGNN} in our experiments, combines the strengths of both GNNs and sequential models, resulting in better performance as shown in our experiments.

\subsection{The Library}

The components discussed in the preceding section have been implemented and made available in an open-source Python library called GraphMuse. This library follows a similar philosophy as PyTorch and PyTorch Geometric, comprising models and graph convolutional blocks, loader pipelines, data pipelines, and related utilities. GraphMuse is built upon and thus requires PyTorch and PyTorch Geometric. The loaders and models provided by GraphMuse are fully compatible with those of PyTorch Geometric. For musical input and output, GraphMuse is compatible with Partitura~\cite{partitura}, a Python library for symbolic music processing, allowing it to work with a variety of input formats such as MusicXML, MEI, Humdrum **kern, and MIDI.

\section{Evaluation}\label{sec:eval}

To evaluate our framework, we perform experiments on two tasks, cadence detection and pitch spelling. We put to the test both the models discussed as well as the sampling process. For pitch spelling, we compare our models to the previous sequential state-of-the-art model, PKSpell~\cite{foscarin2021pkspell} and the GraphSAGE variant of our note-level model. For cadence detection, we compare our models to the previous state-of-the-art model by Karystinaios and Widmer~\cite{karystinaios2022cadence} which is also graph-based and follows a GraphSAGE sampling strategy. For both tasks, we perform ablations by removing the hierarchical elements such as beat and measure nodes and edges, or incorporating hybrid models. This work focuses on the application of the GraphMuse library therefore, a detailed comparison of various input encodings and architectures, as conducted by~\cite{zhang2023symbolic}, is beyond the scope of this paper.

\subsection{Pitch Spelling}\label{subsec:pitch_spelling}

Previous work on Pitch Spelling set the state-of-the-art by using a sequential model~\cite{foscarin2021pkspell}. The task of pitch spelling tackles in parallel key signature estimation and pitch spelling estimation per note, however, the key signature is a global attribute usually set for the whole piece although it can sometimes change midway. The previous architecture uses a GRU encoder for pitch spelling and then infuses the logits together with the latent representation to another GRU layer for the key signature prediction.

For our approach, we use a GNN encoder as described in Section~\ref{subsec:models} followed by two classification heads for key and pitch spelling respectively.
We train and evaluate all models on the ASAP dataset~\cite{peter2023automatic} using a random split with $15\%$ of the data for testing and the $85\%$ for train and validation as described in ~\cite{foscarin2021pkspell}. 


\subsection{Cadence Detection}\label{subsec:cad_model}

For the cadence detection model, we chose to use a modified version of the cadence detection model originally proposed in~\cite{karystinaios2022cadence}. Our considerations were based on a more efficient training process, and the integration of our pipeline possibilities.
The model was expanded to accept a heterogeneous score graph as input, as described in Section~\ref{subsec:graph_creation}. Additionally, we enhanced the model's predictive capabilities from binary (no-cad or PAC) to multiclass cadence prediction, encompassing PAC, IAC, and HC labels. Furthermore, we refined the architecture by incorporating an onset regularization module, which aggregates the latent representations (post-GNN encoder) of all notes occurring at a distinct onset within the score and assigns them to every note sharing that onset. 

In the training phase, the input graph first undergoes processing through the graph encoder. The resulting node embeddings are then grouped based on onset information extracted from the score, and their representations are averaged. Subsequently, embedded SMOTE~\cite{chawla2002smote} is applied to balance the distribution of cadence classes compared to the notes lacking cadence labels in the score. However, during inference, this synthetic oversampling step is omitted. Finally, the oversampled embeddings are fed into a shallow 2-layer MLP classifier to predict the cadence type.

We trained our model with a joined corpus of cadence annotations from the DCML corpora\footnote{\url{https://github.com/DCMLab/dcml_corpora}}, the Bach fugues from the well-tempered clavier Book No.1 ~\cite{giraud2015computational}, the annotated Mozart string quartets~\cite{allegraud2019learning}, and the annotated Haydn string quartets~\cite{sears2018simulating}. Our joined corpus makes for $590,149$ individual notes and $17,188$ cadence annotations. We use 80\% of the data for training and validation and test on 20\% using a random split. Note that these results cannot be directly compared with \cite{karystinaios2022cadence} since we use a different (bigger) dataset and perform multiclass prediction.

\subsection{Experiments}\label{subsec:results}

\subsubsection{Configuration}

The configuration for training pitch spelling graph models with our sampling technique uses a batch size \( B = 300 \), sampling from 300 scores at each step, and target node size \( S = 300 \). For cadence graph models, \( B = 200 \) and \( S = 500 \). All graph models, including GraphSAGE, utilize three heterogeneous SageConv layers with a hidden size of 256 and a dropout of 0.5. Neighbor sampling for each layer fetches up to three neighbors per sampled node per relation. We train all models with the Adam optimizer (learning rate \( 10^{-3} \), weight decay \( 5 \times 10^{-4} \)) on a GTX 1080 Ti. Each experiment is repeated at least four times with different random seeds, and statistical significance testing is performed using the ASO method at a confidence level \( \alpha = 0.05 \)~\cite{dror2019deepsig}~\footnote{For the detailed experiments visit: \url{https://wandb.ai/melkisedeath/GraphMuse}}.

\subsubsection{Results}
Table~\ref{fig:main_results} presents the results of experiments experiments conducted on the two tasks. The metrics used for evaluation are Accuracy (\textit{A}) for pitch spelling and key recognition, and the macro F1 score (\textit{F1}) for cadence detection.  Note that the model employed on the GraphSAGE methods and the model NoteGNN are virtually the same apart from the sampling strategy with which they were trained. 

For the pitch spelling task, we can observe that the actual pitch spelling accuracy (A-Pitch) of all proposed models surpasses both the PKSpell and GraphSAGE methods. Across all models, the MetricalGNN achieves the highest accuracy of $95.6\%$, closely followed by BeatGNN and MeasureGNN with accuracies of $95.1\%$ and $95.4\%$, respectively. These results indicate the benefits of incorporating hierarchical musical elements such as beats and measures. However, it is worth noting that while MetricalGNN achieves the highest accuracy,
it is closely followed by the hybrid model, HybridGNN, which achieves an accuracy of $95.4\%$, suggesting that competitive performance can also be achieved by mixing model types. 

Focusing on the key estimation subtask (A-Key) of pitch spelling we notice that the PKSpell model achieves a very good key accuracy of $69.9\%$, closely followed by the MeasureGNN model and only surpassed by the Hybrid model. We attribute the effectiveness of key detection of a sequential model such as PKSpell to the persistence of the key label across elements of the sequence. Therefore, a hybrid model in this case seems to be able to adapt to the diversity of labels for pitch spelling and uniformity of labels for key estimation. We found our best model to be stochastically dominant over PKSpell with $min\_\epsilon=0.17$. 

In the cadence detection task, we evaluate the results using the macro F1 score to account for the overwhelming presence of non-cadence nodes, as instructed by~\cite{karystinaios2022cadence}. We observe that GraphSAGE, the previously used technique for training, obtains the lowest F1 score and it is surpassed by all the proposed GNN-based models trained with the new sampling method. 

Among our GNN models, BeatGNN and HybridGNN achieve the highest scores of $57.4\%$ and $58.6\%$, respectively, closely followed by MeasureGNN. In this case, the MetricalGNN model surprisingly does not achieve such a good score even though it includes both measure and beat elements. However, it still performs better than the NoteGNN and the GraphSAGE method.


Overall, the results demonstrate the efficacy of GNN-based models trained using our new sampling method. Incorporating hierarchical elements such as beats and measures improves both pitch spelling and cadence detection tasks. Additionally, the hybrid approach of combining GNNs with sequential models produces promising results.

\begin{table}
  \begin{tabular}{l|c|c|c}    
    \multirow{2}{*}{Task} &
      \multicolumn{2}{c}{\textbf{Pitch Spelling}} &      
      \textbf{Cadence} \\
    & A-Pitch & A-Key & F1-Cad \\
    \hline
    PKSpell & $94.8\pm 0.5$ & $\underline{69.9}\pm 1.6$ & - \\
    GraphSAGE & $93.6\pm 0.1$ & $43.3\pm 0.1$ & $53.5\pm 0.8$ \\
    \hline
    NoteGNN & $94.9\pm 0.1$ & $69.3\pm 7.0$ & $55.3\pm 0.9$\\
    BeatGNN & $95.1\pm 0.2$ & $68.7\pm 1.1$ & $\underline{57.4}\pm 1.2$ \\
    MeasureGNN & $\underline{95.4}\pm 0.3$ & $69.5\pm 7.2$ &  $57.0\pm 1.0$ \\ 
    MetricalGNN & $\textbf{95.6}\pm 0.1$ & $64.4\pm 5.3$ & $55.8\pm 0.6$ \\
    HybridGNN & $\underline{95.4}\pm 0.2$ & $\mathbf{72.6}\pm 2.8$ & $\mathbf{58.6}\pm 0.7$\\ 
  \end{tabular}
  \caption{Results on the two tasks, in terms of accuracy (\textit{A}) and F1 score, respectively. Values in bold are the best score per metric; underlined values are the second best. All runs are repeated 4 times. $\pm$ indicates standard deviation.}
  \label{fig:main_results}
\end{table}




\section{Conclusion}


In this paper, we introduced GraphMuse, a framework and Python library for symbolic graph music processing. We designed a specialized sampling process for musical graphs and demonstrated our pipeline's effectiveness through experiments on pitch spelling and cadence detection. Our results show that carefully designed GNN architectures, especially those incorporating hierarchical elements like beats and measures, can lead to better performance. Finally, hybrid models that integrate GNNs with sequential models yield further performance improvements.

Future research will focus on refining GNN-based models in music processing, adding more tasks, and exploring novel architectures. This includes investigating advanced graph convolutional blocks, other sampling techniques, and attention mechanisms to enhance model performance. 


\section{Acknowledgements}
The authors would like to thank Nimrod Varga for his contribution to accelerating graph creation by adapting it to C code. This work is supported by the European Research Council (ERC) under the EU's Horizon 2020 research \& innovation programme, grant agreement No.\ 101019375 (\textit{Whither Music?}), and the Federal State of Upper Austria (LIT AI Lab).

\bibliography{biblio}
\end{document}